\documentclass[twocolumn,aps,prd,amsmath,superscriptaddress]{revtex4-1}
\usepackage{setspace}
\usepackage{color}
\usepackage{fancyhdr}
\usepackage{graphicx}
\usepackage[ansinew]{inputenc}
\usepackage{amssymb}
\usepackage{amsmath}
\usepackage{cancel}
\usepackage{float}
\usepackage{graphicx}
\usepackage{float}
\usepackage{todonotes}
\usepackage{pgfplots}
\pgfplotsset{compat=1.9}
\usepackage{xparse}
\usepackage{tikz}
\usetikzlibrary{trees}
\usetikzlibrary{decorations.pathmorphing}
\usetikzlibrary{decorations.markings} 
\usetikzlibrary{positioning,arrows,patterns}
\usetikzlibrary{calc}

 \tikzset{
    photon/.style={decorate, decoration={snake}, draw=red},
    electron/.style={draw=blue, postaction={decorate},
        decoration={markings,mark=at position .55 with {\arrow[draw=blue]{>}}}},
    electron2/.style={draw=blue, postaction={decorate},
        decoration={markings,mark=at position .55 with {\arrow[draw=blue]{<}}}},    
    quark/.style={draw=blue, postaction={decorate},
        decoration={markings,mark=at position .55 with {\arrow[draw=blue]{>}}}},
    quark2/.style={draw=blue, postaction={decorate},
        decoration={markings,mark=at position .55 with {\arrow[draw=blue]{<}}}},    
    gluon/.style={decorate, draw=green, decoration={coil,amplitude=4pt, segment length=5pt}}, 
    fermion/.style={draw=black, postaction={decorate},decoration={markings,mark=at position .55 with {\arrow{>}}}},
    vertex/.style={draw,shape=circle,fill=black,minimum size=3pt,inner sep=0pt},
}
\NewDocumentCommand\semiloop{O{black}mmmO{}O{above}}
{%
\draw[#1] let \p1 = ($(#3)-(#2)$) in (#3) arc (#4:({#4+180}):({0.5*veclen(\x1,\y1)})node[midway, #6] {#5};)
}

\begin{document}
\title{Effects of Renormalizing the chiral SU(2) Quark-Meson-Model}
\author{Andreas Zacchi}
\email{zacchi@astro.uni-frankfurt.de}
\affiliation{Institut f\"ur Theoretische Physik, Goethe Universit\"at Frankfurt, 
Max von Laue Strasse 1, D-60438 Frankfurt, Germany}

\author{J\"urgen Schaffner-Bielich}
\email{schaffner@astro.uni-frankfurt.de}
\affiliation{Institut f\"ur Theoretische Physik, Goethe Universit\"at Frankfurt, 
Max von Laue Strasse 1, D-60438 Frankfurt, Germany}
\date{\today}
   \begin{abstract}
   
We investigate the restoration of chiral symmetry at finite temperature in the SU(2) quark meson model where the mean field approximation is compared to the renormalized version for quarks and mesons. 
In a combined approach at finite temperature all the renormalized versions 
show a crossover transition. 
The inclusion of different renormalization scales leave the order parameter and the mass spectra 
nearly untouched, but strongly influence the thermodynamics at low temperatures and around the phase transition. We find unphysical results for the renormalized version of mesons and the combined one.
   \end{abstract}
        \maketitle
  \section{Introduction}
Since QCD is non-perturbative in the low energy regime, effective 
theories and models based on the QCD Lagrangian and its properties have 
to be utilized \cite{Toimela:1984xy,Mocsy:2004ab,Braaten96,Fraga:2004gz}.
The QCD Lagrangian possesses an exact color- and flavor symmetry 
for $N_f$ massless quark flavours
\cite{Gasi69,Koch:1997ei,Karsch:2000kv,Karsch:2000hh,Zschiesche_review,Parganlija:2012gv,Parganlija:2012fy}  
and chiral symmetry controls the hadronic interactions 
in the low energy regime \cite{Gavin94a,Chandra99}. 
At high temperatures or densities chiral symmetry is 
expected to be restored \cite{Kirzhnits:1972ut,Pisarski:1983ms}.
In general, the interaction can be 
modeled by the exchange of scalar-, pseudoscalar- and vector 
mesons \cite{Kaym85}. 
If one adopts the linear sigma model \cite{GellMann:1960np,Gell64} for quark interactions, 
it is referred to as the chiral \textit{Quark Meson} model 
\cite{Kogut83,Koch95,Schaefer:2004en,Parganlija:2010fz,Parganlija:2012gv}, which 
is well studied 
\cite{Gupta:2011ez,Herbst:2013ufa,Stiele:2013pma,Lenaghan:1999si,Lenaghan:2000ey,Grahl:2011yk,Seel:2011ju}.
Its advantage in comparison to other chiral effective models like the 
Nambu-Jona-Lasinio model 
\cite{Hara66,Bernard:1995hm,Schertler:1999xn,Buballa:2003qv} 
lies in its renormalizability. Renormalizability takes into 
account the contribution of vacuum fluctuations \cite{Skokov:2010sf,Gupta:2011ez,Chatterjee:2011jd}.
Works which included the vacuum term by using the renormalization group flow equations 
focussed in particular on the neighborhood of critical points 
\cite{Bailin:1984ak,Berges:1997eu,Schaefer:2004en,Mocsy:2004ab}.\\
In this article we study quarks, by using a chiral SU(2) Quark Meson model within the path integral formalism, and mesons, which are examined within the 2PI formalism, within a combined approach. 
We investigate this approach also in the mean field approximation and consider the vacuum term contribution, which  depends on a renormalization scale resulting from the inclusion of the meson fields.

Besides the order parameter and the masses of the sigma and the pion, we study 
thermodynamical quantities.
In all cases studied, the masses of the pion and the sigma meson 
start to be degenerate around the phase transition, which is defined by the order parameter. 
The impact of the meson contribution on the order parameter and mass is 
comparatively small, whereas thermodynamic quantities are strongly influenced. 
At low temperatures the impact of the mesonic contribution 
is substantial within the combined approach. 
 In our approach we vary 
the mass of the sigma meson in the range $500 \leq m_{\sigma}^{vac} \leq 900$~MeV. 
For the standard value of $m_{\sigma}=550$~MeV we find a smooth chiral crossover phase transition around the critical temperature 
$T_c\simeq 155$~MeV \cite{Mocsy:2004ab,Seel:2011ju}. 
We compare our studies for the quark fields with 
works from refs.~\cite{Gupta:2011ez,Herbst:2013ufa,Stiele:2013pma},  
and for the mesonic fields with works from 
refs.~\cite{Lenaghan:1999si,Lenaghan:2000ey,Grahl:2011yk,Seel:2011ju}.
In the combined approach we compare our results with 
the work from ref.~\cite{Mocsy:2004ab}, 
who derive an effective action for the meson fields and linearize 
it around the ground state. \\
We find that the renormalization scale cancels when considering the SU(2) quark-meson model for the quark fields, and the inclusion of the vacuum term shifts the phase transition to larger temperatures. The combined model is dependent on the renormalization scales. Hence, a combined model for quarks and mesons is only acceptable in the mean field approximation.
  \section{General considerations}
Before going into more details, 
we briefly sketch a general consideration to show that the approach used is thermodynamically consistent. We thank Dirk Rischke for pointing this out to us.
A general ansatz for the effective action $\Gamma[\phi,G,Q]$ according to 
\cite{Cornwall:1974vz,Lenaghan:1999si,Mandanici:2003vt,Mocsy:2004ab} is
\begin{eqnarray}\label{wholeformalism_inprops}
 \Gamma[\phi,G,Q]&=&I[\phi]\\ \nonumber
                 &-&\frac{1}{2}\rm{Tr}\left(\ln G^{-1}\right)-\frac{1}{2}\rm{Tr}\left(D^{-1}G-1\right)\\ \nonumber
                 &+&\rm{Tr}\left(\ln Q^{-1}\right)+\rm{Tr}\left(S^{-1}Q-1\right)\\ \nonumber
                 &+&\Gamma_2[\phi,G,Q]
\end{eqnarray}
where $\phi$ represents the fields involved, $I[\phi]$ is the classical action or the tree-level potential. 
G is the full propagator and $D^{-1}$ the inverse tree level propagator for the mesons. Q is the full propagator and $S^{-1}$  
the inverse tree level propagator for the quarks. $\Gamma_2[\phi,G,Q]$ is the contribution from the two-particle irreducible diagrams, which in our case only depends on the fields and the full propagator of the mesons, i.e. $\Gamma_2[\phi,G]$, see also Figure \ref{two_pi_ei_luhps}.\\
In the absence of sources the stationary conditions determine the vacuum expectation values of $\phi$. They read
\begin{eqnarray}\nonumber
 \frac{\delta \Gamma[\phi,G,Q]}{\delta \phi}&=&\frac{\delta I[\phi]}{\delta \phi}-\frac{1}{2}\rm{Tr}\left(\frac{\delta D^{-1}}{\delta \phi}G\right)\\
 &+&\rm{Tr}\left(\frac{\delta S^{-1}}{\delta \phi}Q\right)+\frac{\delta \Gamma_2[\phi,G]}{\delta \phi}=0\\
  \frac{\delta \Gamma[\phi,G,Q]}{\delta G}&=&-\frac{1}{2}D^{-1}+\frac{1}{2}G^{-1}+\frac{\delta \Gamma_2[\phi,G]}{\delta G}=0\\
   \frac{\delta \Gamma[\phi,G,Q]}{\delta Q}&=&-G^{-1}+S^{-1}=0
\end{eqnarray}
Since no contribution from $\Gamma_2[\phi,G]$ to the stationary conditions 
occurs for the quark propagator Q, no 
diagrams containing a quark propagator within a meson loop appear 
within our approach. Hence it is justified to evaluate the potentials 
independently and the respective gap equations in the combined approach are 
consequently additive.\\
In the following we briefly sketch the derivation of the individual approaches to finally combine them.
     \section{Quark-Quark Interaction}\label{qq_i_actins}
A Lagrangian with $N_f=2$ respecting quark fields may be written as 
\cite{Kogut83,Koch95,Lenaghan:1999si,Lenaghan:2000ey}
\begin{eqnarray} \label{eq:qq_lagrangian}
 \mathcal{L}&=&\mathcal{L}_q + \mathcal{L}_m - U(\sigma,\vec{\pi})\\ \label{eq:qq_lagrangian1}
 &=&\bar{\Psi}\left( i \cancel{\partial} -g(\sigma+i\gamma_5\vec{\tau}
 \cdot\vec{\pi})\right)\Psi \\ \label{eq:qq_lagrangian2}
 &+& \frac{1}{2}\left(\partial_{\mu}\sigma\partial^{\mu}\sigma + 
 \partial_{\mu}\vec{\pi}\partial^{\mu}\vec{\pi}\right) - U(\sigma,\vec{\pi})
\end{eqnarray}
where $g={m_{q,vac}}/{f_{\pi}}$ is a Yukawa type coupling 
to the quark spinors $\Psi$. Here 
$m_{q,vac}$ is the constituent quark mass chosen to be 
300 MeV and $f_{\pi}=92.4$~MeV the pion decay constant \cite{Parganlija:2012fy}. 
$U(\sigma,\vec{\pi})$ is the tree level potential and given as
\begin{equation} \label{eq:qq_tree_level_pot}
U(\sigma,\vec{\pi})=\frac{\lambda}{4}(\sigma+\vec{\pi})^4+\frac{m^2}{2}(\sigma+\vec{\pi})^2-H\sigma
\end{equation} 
with the coupling $\lambda$ and the mass term $m=-\lambda v^2$. The term $H$ breaks chiral 
symmetry explicitly and is therefore responsible for the non-vanishing mass 
of the pion \cite{Sche71b,Kogut83,Vafa:1983tf,Koch95,Torn97}.
The grand canonical potential is commonly derived with the path integral 
formalism  \cite{Bochkarev:1995gi,Berges:1997eu,Scavenius:2000qd,Mocsy:2004ab,Schaefer:2008hk,Zacchi:2015lwa} and reads  
\begin{eqnarray} \label{eq:qq_granpotall}
 \Omega_{\bar{q}q} &=& U(\sigma,\vec{\pi}) + \Omega_{\bar{q}q}^{th} + \Omega_{\bar{q}q}^{vac}\\  \label{tree_beard}
 &=&\frac{\lambda}{4}(\sigma+\vec{\pi})^4+\frac{m^2}{2}(\sigma+\vec{\pi})^2-H\sigma\\ \label{eq:qq_granpotall_therm}
 &-& {N_f N_c T}\int_0^{\infty}\frac{dk^3}{\left({2\pi}^3\right)}\left[\ln \left(1+ e^{-\beta(E_k \pm \mu_f)}\right)\right]\\ \label{eq:qq_granpotall_vac}
 &-& {N_f N_c T}\int_0^{\infty}\frac{dk^3}{\left({2\pi}^3\right)}\left(\frac{E}{T}\right)
\end{eqnarray}
Here $N_c=3$, the single particle energy 
\begin{equation}\label{eq:sipaen}
E_k=\sqrt{k^2+\tilde{m}_f^2} \quad\hbox{with}\,\, \tilde{m}_f=g \sqrt{\sigma^2+\vec{\pi}^2}
\end{equation}
as the effective mass, and $\mu_f$ 
as the flavour dependent quark chemical potential, have been introduced. 
The term of line~(\ref{eq:qq_granpotall_vac}) represents the contribution due to vacuum fluctuations. 
Solutions are then obtained by solving  
\begin{equation}\label{condensate_eq}
 \frac{\partial \Omega_{\bar{q}q}}{\partial \sigma}\overset{!}{=}0, \hspace{.2cm} 
 \frac{\partial^2\Omega_{\bar{q}q}}{\partial\sigma^2}=m_{\sigma} \quad\hbox{and}\,\,
 \frac{\partial^2\Omega_{\bar{q}q}}{\partial\vec{\pi}^2}=m_{\vec{\pi}}
\end{equation}
also known as gap equations.\\
The vacuum parameters can be found in Tab.~\ref{wirsindallebluna}.
      \subsection{Regularization for the quark fields}\label{sec:qq_vac}
Taking into account vacuum fluctuations needs regularization schemes 
\cite{Lenaghan:1999si,Lenaghan:2000ey,Mocsy:2004ab,Skokov:2010sf}.
To regularize the divergencies we use dimensional regularization.\\
The vacuum term in eq.~(\ref{eq:qq_granpotall}) (eq.~(\ref{eq:qq_granpotall_vac})), 
is, to lowest order just the one-loop
effective potential at zero temperature and reads in $d=3-2\epsilon$ dimensions, where 
$\lim\epsilon\rightarrow 0$, regularized \cite{Skokov:2010sf}
\begin{equation}
 \Omega_{\bar{q}q}^{vac}=\frac{N_c N_f}{16\pi^2}\tilde{m}_f^4\left[\frac{1}{\epsilon}-\frac{1}{2}\left[ -3+2\gamma+4\rm{ln}\left(\frac{\tilde{m}_f}{2\sqrt{\pi}\Lambda}  \right)\right]\right]
\end{equation}
Here $\gamma$ is the Euler-Mascheroni constant and $\Lambda$ an arbitrary renormalization 
scale parameter. 
To renormalize the thermodynamic potential an appropriate counter 
term $\delta \mathcal{L}$ needs to be introduced to the Lagrangian \cite{Skokov:2010sf}. 
The minimal substraction ($\overline{MS}$) scheme allows for
\begin{equation}
 \delta \mathcal{L}=\frac{N_c N_f}{16\pi^2}\tilde{m}_f^4\left[\frac{1}{\epsilon}-\frac{1}{2}\left[ -3+2\gamma-4\rm{ln}\left(2\sqrt{\pi}  \right)\right]\right]
\end{equation}
and the renormalized vacuum contribution becomes 
\begin{equation} \label{eq:renorn_th_pot}
\Omega_{\bar{q}q}^{vac} \quad \rightarrow \quad \Omega_{\bar{q}q}^{dr}=-\frac{N_c N_f}{8\pi^2}\tilde{m}_f^4 \ln\left(\frac{\tilde{m}_f}{\Lambda}\right) 
\end{equation}

The vacuum contributions to the gap equations, eqs.~(\ref{condensate_eq}), due to eq.~(\ref{eq:renorn_th_pot}) are
\begin{eqnarray}
\frac{\partial\Omega_{\bar{q}q}^{dr}}{\partial\sigma}&=&-\frac{N_c N_f g^4 \sigma^3}{8\pi^2}\left[ 1+4\ln\left( \frac{\sigma}{f_{\pi}}\right) \right]\\
\frac{\partial^2\Omega_{\bar{q}q}^{dr}}{\partial\sigma^2}&=&-\frac{N_c N_f g^4 \sigma^2}{8\pi^2}\left[ 7+12\ln\left( \frac{\sigma}{f_{\pi}}\right) \right]\\
\frac{\partial^2\Omega_{\bar{q}q}^{dr}}{\partial\vec{\pi}^2}&=&-\frac{N_c N_f g^4 \sigma^2}{8\pi^2}\left[ 1+4\ln\left( \frac{\sigma}{f_{\pi}}\right) \right]
\end{eqnarray}
Note that $\Lambda$ cancels in the determination of the vacuum parameters (case $Q_{th+vac}$ in Tab.~\ref{wirsindallebluna}) and hence the grand canonical potential is also independent on the 
choice of $\Lambda$. This is also the case for an SU(3) approach \cite{Chatterjee:2011jd,Gupta:2011ez,Tiwari:2013pg}.

  \section{The 2PI formalism}\label{mesonic_inter}

At finite temperature perturbative expansion in powers of the coupling constant breaks
down due to infrared divergencies, and an approach for the mesonic fields via the path integral formalism 
leads to difficulties, because at low momentum spontaneous symmetry breaking for 
instance leads to quasi particle exitations with imaginary 
energies  \cite{Lenaghan:1999si,Lenaghan:2000ey,Seel:2011ju}.\\ 
These difficulties can be circumvented utilizing 
the Cornwall-Jackiw-Toumboulis (CJT) \cite{Cornwall:1974vz}, or more 
commonly, 2PI formalism, 
which is understood as a relativistic generalization of the Luttinger 
Ward formalism \cite{Pilaftsis:2013xna,Dupuis:2013vda}. 
The 2PI formalism can be viewed as a prescription for computing the effective action of a 
theory, where the stationary conditions are the Greens functions and 
the effective action corresponds to the effective potential \cite{Cornwall:1974vz}.
However, the in-medium masses of the $\sigma$- and the $\pi$-meson can then be solved 
self-consistently \cite{Lenaghan:1999si,Lenaghan:2000ey}. 
The grand canonical potential can be derived via the generating functional 
for the respective Greens functions \cite{Cornwall:1974vz}, which, in the presence 
of the two sources J and K, is given as
\begin{equation}
 Z[J,K]=e^{\mathcal{W}[J,K]}=\int\mathcal{D}\phi e^{(\phi J+\frac{1}{2}\phi K \phi+I[\phi])}
\end{equation}
with $\mathcal{W}[J,K]$ as the generating functional for the connected Greens functions.
$I[\phi]=\int_x \mathcal{L}$ is the classical action with $\mathcal{L}=\mathcal{L}_m+U(\sigma,\vec{\pi})$ 
from eq.~(\ref{eq:qq_lagrangian2}). 
Throughout this article we stick to the shorthand notation  
\begin{equation}
 \int_x F(x)=\int_{0}^ {\beta} d\tau \int d^3 \vec{k}F(\tau,\vec{k})
\end{equation}
for the corresponding integrals, where $\mathcal{F}$ is the appropriate distribution function.\\  

The effective action according to \cite{Cornwall:1974vz} is 
\begin{eqnarray}
 \Gamma[\bar{\phi},G]&=&I[\bar{\phi}] - \frac{1}{2} Tr (D^{-1}G-1) \\ \nonumber
                     &-& \frac{1}{2} \rm{Tr}(\ln G^{-1})+\Gamma_2[\bar{\phi},G]
\end{eqnarray}
with $D^{-1}$ as the inverse tree level propagator and G as full propagator. 
$\Gamma_2[\bar{\phi},G]$ represents the sum of all two particle irreducible diagrams, 
see fig.~\ref{two_pi_ei_luhps}, where all lines represent full propagators G. 
In momentum space 
\begin{equation}\label{D_invers}
D^{-1}(k,\bar{\phi})=-k^2+U''(\bar{\phi}) 
\end{equation}
and the full propagator is 
\begin{equation}\label{eq:g_normahl}
G_{\sigma,\pi}(k)=\frac{1}{-k^2+\bar{m}_{\sigma,\pi}^2}
\end{equation}
For constant fields $\bar{\phi}(x)=\bar{\phi}$ and homogenous systems, 
the effective potential is \cite{Cornwall:1974vz,Lenaghan:1999si,Lenaghan:2000ey,Mandanici:2003vt}
\begin{eqnarray}\nonumber
 \Omega[\bar{\phi},G]&=&U(\bar{\phi})+\frac{1}{2}\int_k \ln G^{-1}(k)\\ \label{eq:2pi_potential}
                     &+&\frac{1}{2} \int_k \left[D^{-1}(k,\bar{\phi})G(k)-1\right]+\Omega_2
\end{eqnarray}
Here $\Omega_2 \equiv -\rm{T} \cdot \Gamma_2[\bar{\phi},G]/V$, V being the 3-volume of the system. 
The 2PI potential reads 
\begin{eqnarray}\label{eq:2pi_potential_O}
\Omega_{2PI}(\phi,G_{\sigma,\pi})&=&\frac{1}{2}m^2\phi^2+\frac{1}{4}\lambda \phi^4 - H\phi \\ \nonumber
&+&\frac{1}{2}\int_k\left[\rm{ln}G_{\sigma}^{-1}(k)+D_{\sigma}^{-1}(k,\phi)G_{\sigma}(k)-1\right]\\ \nonumber
&+&\frac{3}{2}\int_k\left[\rm{ln}G_{\pi}^{-1}(k)+D_{\pi}^{-1}(k,\phi)G_{\pi}(k)-1\right]\\ \nonumber
&+&\Omega_2
\end{eqnarray}
with the two loop contribution to the potential 
\begin{eqnarray}\label{eq.omegazwoh}
\Omega_2&=&\frac{3 \lambda}{4}\left[\int_k G_{\sigma}(k)\right]^2 + \frac{15 \lambda}{4}\left[\int_k G_{\pi}(k)\right]^2\\ \nonumber 
&+& \frac{3 \lambda}{2}\left[\int_k G_{\sigma}(k)\right]\left[\int_k G_{\pi}(k)\right]  \nonumber 
\end{eqnarray}
The respective diagrammatic expressions for the potential from eq.~(\ref{eq:2pi_potential_O}) 
are shown in Figs.~\ref{one_pi_ei_luhps} and \ref{two_pi_ei_luhps}.\\
The gap equations obtained via eqs.~(\ref{condensate_eq}) for the meson fields read
\begin{eqnarray}\label{eq:gap_cjt}
H&=&\phi \left[m^2+\lambda\left(\phi^2+3F(\bar{m}_{\sigma},T)+3F(\bar{m}_{\pi},T)\right)\right] \\ \label{eq:gap2}
\bar{m}_{\sigma}&=&m^2+\lambda\left[3\phi^2+3F(\bar{m}_{\sigma},T)+3F(\bar{m}_{\pi},T)\right]\\ \label{eq:gap3}
\bar{m}_{\pi}&=&m^2+\lambda\left[\phi^2+F(\bar{m}_{\sigma},T)+5F(\bar{m}_{\pi},T)\right]
\end{eqnarray}
Herein the function 
\begin{eqnarray} \label{temp_dep_fnct_F}
	&F(\bar{m}_{\sigma,\pi},T)&=F_{T}(\bar{m}_{\sigma,\pi},T)+F_{vac}(\bar{m}_{\sigma,\pi},T) \\ \nonumber
	&=&\int \frac{d^3\vec{k}}{(2\pi)^3}\frac{1}{\sqrt{\vec{k}^2+\bar{m}_{\sigma,\pi}^2}}\cdot\left[ {\frac{1}{e^{\beta {\sqrt{\vec{k}^2+\bar{m}_{\sigma,\pi}^2}}}-1}} +\frac{1}{2}\right]
\end{eqnarray}
displays the temperature dependence including the vacuum contribution \cite{Lenaghan:1999si}. 
For more details on the calculation see \cite{Cornwall:1974vz,Lenaghan:1999si,Lenaghan:2000ey,Mandanici:2003vt}.
The vacuum parameters are listed in Tab.~\ref{wirsindallebluna}. 
\begin{figure}[H]
\center 
\begin{tikzpicture}
\filldraw[color=black!80, fill=black!.01, thick](0,0.75) circle (0.75);
\filldraw [black] (0,0) circle (3pt);
   \draw[-] [black, thick] (0,0) -- (0.75,-0.75); 
   \draw[-] [black, thick] (0,0) -- (-0.75,-0.75); 
\filldraw[color=black!80, fill=black!.01, thick](3,0.75) circle (0.75);
\filldraw [black] (3,0) circle (3pt);
   \draw[-] [black, thick] (3,0) -- (3.75,-0.75); 
   \draw[-] [black, thick] (3,0) -- (2.25,-0.75);    
\filldraw [black] (3,1.5) circle (3pt);   
   \draw[-] [black, thick] (3,1.5) -- (3.75,2.25); 
   \draw[-] [black, thick] (3,1.5) -- (2.25,2.25);    
\filldraw[color=black!80, fill=black!.01, thick](6,0.75) circle (0.75);
\filldraw [black] (6,0) circle (3pt);
   \draw[-] [black, thick] (6,0) -- (6.75,-0.75); 
    \draw[-] [black, thick] (6,0) -- (5.25,-0.75);
     \filldraw [black] (5.3,1.0) circle (3pt);   
 \draw[-] [black, thick] (5.3,1.0) -- (5.3,2); 
    \draw[-] [black, thick] (5.3,1.0) -- (4.3,1);
  \filldraw [black] (6.7,1.0) circle (3pt);
     \draw[-] [black, thick] (6.7,1.0) -- (6.7,2); 
    \draw[-] [black, thick] (6.7,1.0) -- (7.7,1);
\coordinate[label=right:$+$] (A) at (1.2,0.55);
 \coordinate[label=right:$+$] (A) at (4.2,0.55);
 \coordinate[label=right:$+ ...$] (A) at (7.2,0.55);
\end{tikzpicture}
 \caption{\textit{The 1-PI loops contributing to the effective potential in eq.~(\ref{eq:2pi_potential}), i.e. eq.~(\ref{eq:2pi_potential_O}) without $\Omega_2$ from eq.~(\ref{eq.omegazwoh}).}}
\label{one_pi_ei_luhps}
\end{figure}
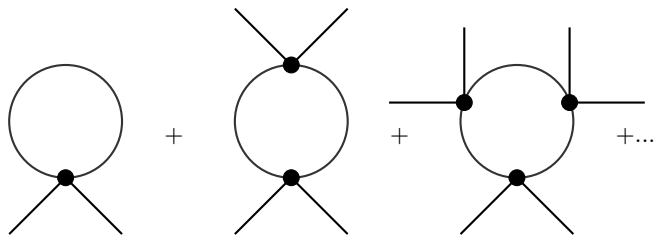
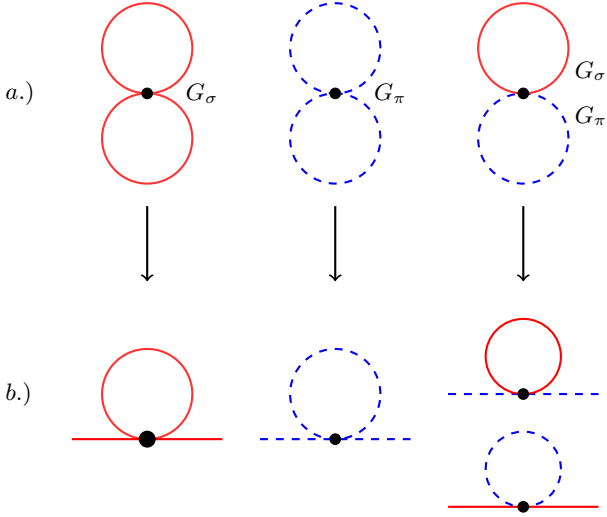
\begin{figure}[H]
\center 
\begin{tikzpicture}
\coordinate[label=right:$a.)$] (A) at (-2,0);
\coordinate[label=right:$b.)$] (B) at (-2,-4);
\filldraw[color=red!80, fill=red!.01, thick](0,.6) circle (0.6);
\filldraw[color=red!80, fill=red!.01, thick](0,-.6) circle (0.6);
\filldraw [black] (0,0) circle (2pt);
\draw[blue,dashed, thick] (2.5,.6) circle (.6);
\draw[blue,dashed, thick] (2.5,-.6) circle (.6);
\filldraw [black] (2.5,0) circle (2pt);
\draw[blue,dashed, thick] (5,-.6) circle (.6);
\filldraw[color=red!80, fill=red!.01, thick](5,.6) circle (0.6);
\filldraw [black] (5,0) circle (2pt);
\coordinate[label=right:$G_{\sigma}$] (A) at (0.4,0);
\coordinate[label=right:$G_{\pi}$] (A) at (2.9,0);
\coordinate[label=right:$G_{\sigma}$] (A) at (5.57,0.3);
\coordinate[label=right:$G_{\pi}$] (A) at (5.57,-0.3);
   \draw[->] [black, thick] (0,-1.5) -- (0,-2.5); 
   \draw[->] [black, thick] (2.5,-1.5) -- (2.5,-2.5); 
   \draw[->] [black, thick] (5,-1.5) -- (5,-2.5);
\filldraw[color=red!80, fill=red!.01, thick](0,-4) circle (.6);
\draw[-] [red, thick] (-1,-4.6) -- (1,-4.6); 
\filldraw [black] (0,-4.6) circle (3pt);
\draw[blue,dashed, thick] (2.5,-4) circle (.6);
\draw[-] [blue,dashed, thick] (1.5,-4.6) -- (3.5,-4.6); 
\filldraw [black] (2.5,-4.6) circle (2pt);
\draw[-] [red, thick] (4,-5.5) -- (6,-5.5); 
\draw[blue,dashed, thick] (5,-5) circle (.5);
\draw[-] [blue,dashed, thick] (4,-4) -- (6,-4); 
\draw[red, thick] (5,-3.5) circle (.5);
\filldraw [black] (5,-5.5) circle (2pt);
\filldraw [black] (5,-4.0) circle (2pt);
\end{tikzpicture}
 \caption{\textit{ a.) The two loop Hartree contributions, eq.~(\ref{eq.omegazwoh}), to the CJT effective potential ($\Omega_2$). The full red line corresponds to $G_{\sigma}$, whereas the dashed blue line corresponds to $G_{\pi}$. The right-most diagram stands for the last term in eq.~(\ref{eq:2pi_potential_O}). 
 b.) the tadpole contribution to the self energy, obtained by cutting a line.}}
\label{two_pi_ei_luhps}
\end{figure}

       \subsection{Regularization for the meson fields}\label{mesons_vacuum}
%
We use the dimensional regularization 
procedure for meson fields \cite{vanHees:2001ik}. 
Whereas for the quark fields we added a counter term to 
the Lagrangian, for the meson fields it is sufficient 
to just add a correction to the mass term, $\delta m$, since no higher order diagrams are considered. 
%
The correction to the naked mass is calculated to be \cite{Lenaghan:1999si,Grahl:2011yk} 
\begin{equation}
 \delta m^2=-\frac{\lambda m^2}{16\pi^2 \epsilon}-\frac{\lambda m^2}{32\pi^2}\ln\left(\frac{4\pi\mu^2 e}{m^2 e^ {\gamma}} \right)+ \mathcal{O}(\epsilon^2)
\end{equation}
Here $\mu$ plays the role of $\Lambda$ from the quark fields, i.e. is an arbitrary renormalization scale parameter.

The procedure is equivalent to the one for the quark fields \cite{Mandanici:2003vt}, utilizing the $\overline{MS}$ scheme. 
The renormalized vacuum contribution from eq.~(\ref{temp_dep_fnct_F}) finally reads  
\begin{eqnarray} \label{eq:vdepp}
F_{vac}(\bar{m}_{\sigma,\pi})&=&\int \frac{d^3\vec{k}}{(2\pi)^3}\frac{1}{2 \sqrt{\vec{k}^2+\bar{m}_{\sigma,\pi}^2}} \\ \nonumber
&=& -\frac{\bar{m}_{\sigma,\pi}}{16\pi^2}\left[1+\rm{ln}\left(\frac{\mu^2}{\bar{m}_{\sigma,\pi}}\right)\right]\equiv F_{dr}(\bar{m}_{\sigma,\pi})
\end{eqnarray}
Again, the vacuum parameters are given in Tab.~\ref{wirsindallebluna}. 
  \section{Combining interactions between Quarks and Mesons}\label{ciwqam}
Since the grand canonical potential is an intensive quantity, it is additive, 
and so are the respective gap equations of the corresponding sectors, 
obtained in each case with eqs.~(\ref{condensate_eq}).
This section now combines both approaches to an unified set of equations. 
Firstly we will treat the thermal contributions only, 
whereas in the following we include the vacuum fluctuations from the quark fields. 
The potential is a sum of the independent potentials
\begin{equation}
 \Omega_{QAM}^{th}=\Omega_{\bar{q}q}^{th}+\Omega_{2PI}(\phi,G_{\sigma,\pi})
\end{equation}
Here $\Omega_{QAM}^{th}$ is the thermal part of the combined grand canonical 
potential of quarks and mesons (QAM).
      \subsection{Regularization for the combined approach}
As mentioned above, all relevant quantities are additive, and so are the vacuum contributions. 
Hence there is no need to regularize and renormalize anew. Both 
equations for the divergent vacuum contributions, eqs.~(\ref{eq:renorn_th_pot}) 
and (\ref{eq:vdepp}), can be merged into a single set of gap equations. 
The potential is the sum of the independent potentials, i.e. eqs.~(\ref{eq:qq_granpotall}) 
and (\ref{eq:2pi_potential_O}). The tree level potential, eq.~(\ref{eq:qq_tree_level_pot}), appears only once.
\begin{equation}
 \Omega_{QAM}=\Omega_{\bar{q}q}^{th}+\Omega_{\bar{q}q}^{dr}+\Omega_{2PI}(\phi,G_{\sigma,\pi})
\end{equation}
The vacuum parameters $\lambda$, $m^2$ and H, obtained by solving 
eqs.~(\ref{condensate_eq}) are determined to be
\begin{eqnarray}\label{lambda_qam}
 \lambda&=&\frac{m_{\sigma}^2+m_{\pi}^2+\frac{N_c N_f}{8\pi^2}g^4\sigma^2\left[6+8\ln\left(\frac{g\sigma}{\Lambda} \right)\right]}{2\left(F_{dr}(\bar{m}_{\sigma})-F_{dr}(\bar{m}_{\pi})+\sigma^2\right)}\\ \nonumber
 m^2&=&\frac{N_c N_f}{8\pi^2}g^4\sigma^2\left[7+12\ln\left(\frac{g\sigma}{\Lambda} \right)\right]\\ \label{mquadraht_qam}
 &-&3\lambda\left(F_{dr}(\bar{m}_{\sigma})+F_{dr}(\bar{m}_{\pi}) \right)+m_{\sigma}^2-3\lambda\sigma^2\\ \nonumber
 H&=&-\frac{N_c N_f}{8\pi^2}g^4\sigma^3\left[1+4\ln\left(\frac{g\sigma}{\Lambda} \right)\right]\\ \label{H_qam_quasimodo}
 &+&3\lambda\sigma\left(F_{dr}(\bar{m}_{\sigma})+F_{dr}(\bar{m}_{\pi})\right)+\sigma(m^2+\lambda\sigma^2)
\end{eqnarray}   
and the corresponding gap equations read
\begin{eqnarray}\label{gaps_at_qam_cops_at_fart_pub}
 \frac{\partial\Omega_{QAM}}{\partial\sigma}&=&-\frac{N_c N_f}{8\pi^2}g^4\sigma^3\left[1+4\ln\left(\frac{g\sigma}{\Lambda} \right)\right]\\ \nonumber
 &+&3\lambda\sigma\left(F(\bar{m}_{\sigma})+F(\bar{m}_{\pi})\right)+m^2\sigma+\lambda\sigma^3=H\\ \label{eq:qam_gaap2}
 \frac{\partial^2\Omega_{QAM}}{\partial\sigma^2}&=&-\frac{N_c N_f}{8\pi^2}g^4\sigma^2\left[7+12\ln\left(\frac{g\sigma}{\Lambda} \right)\right]\\ \nonumber
 &+&3\lambda\left(F(\bar{m}_{\sigma})+F(\bar{m}_{\pi})\right)+m^2+3\lambda\sigma^2=m_{\sigma}^2\\ \label{piqaum}
 \frac{\partial^2\Omega_{QAM}}{\partial\pi^2}&=&-\frac{N_c N_f}{8\pi^2}g^4\sigma^2\left[1+4\ln\left(\frac{g\sigma}{\Lambda} \right)\right]\\ \nonumber
 &+&\lambda\left(F(\bar{m}_{\sigma})+F(\bar{m}_{\pi})\right)+m^2+\lambda\sigma^2=m_{\pi}^2
 \end{eqnarray} 
Unfortunately these equations leave us with the possibility of having 
two renormalization scales, one from the quark-quark contribution, $\Lambda$, and 
one hidden in $F(\bar{m}_{\sigma\pi})$, namely $\mu$ (see eq.~(\ref{eq:vdepp})). 
The vacuum parameters are listed in Tab.~\ref{wirsindallebluna}. 

\section{Results for the renormalized quark fields}\label{resultsinthequarkssector}
\begin{figure}[H]
\center
\includegraphics[width=1.0\columnwidth]{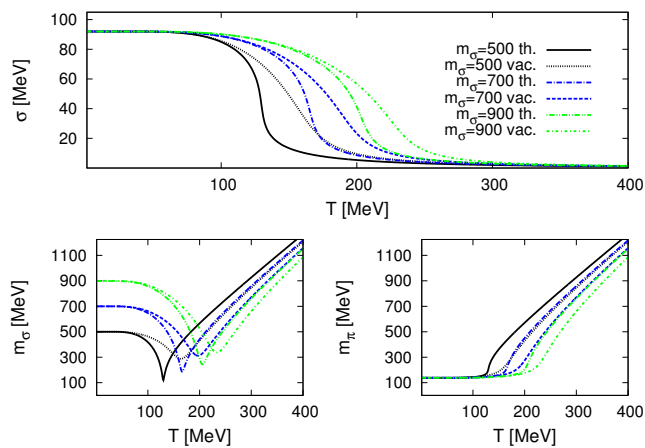}
\caption{\textit{The $\sigma$ condensate as a function of temperature for zero chemical potential without (denoted as ``th.'') and with vacuum contribution (denoted as ``vac.'') for three different values of the vacuo sigma meson mass $m_{\sigma}^{vac}$ shown in the upper figure. The lower figures shows the in-medium masses of the sigma and the pi.}}
\label{fig:felderli_quarkse}
\end{figure}

The upper part of Figure \ref{fig:felderli_quarkse} shows the order 
parameter $\sigma$ as a function of 
the temperature for three different vacuum sigma meson masses 
$m_{\sigma}^{vac}$, neglecting (denoted in the figures as ``th.'') 
and including (denoted in the figures as ``vac.'') the vacuum term of the quarks. This corresponds to the cases $Q_{th}$ and $Q_{th+vac}$ in Tab.~\ref{wirsindallebluna}.\\
We find that with increasing vacuum sigma meson mass $m_{\sigma}^{vac}$ the phase
transition in the thermal case is shifted to higher 
temperatures and becomes slightly more crossover like, 
whereas smaller values of $m_{\sigma}^{vac}$ lead to a behaviour close to a first order
phase transition, which is not achieved even for our lowest choice of 
$m_{\sigma}^{vac}=500$~MeV. 
The curves containing the vacuum contribution show the same behaviour, 
only the trends are noticeable more crossover like, 
and hence shifted to higher transition temperatures with increasing values of $m_{\sigma}^{vac}$.
The behaviour of the order parameter $\sigma$  
can be translated to the behaviour of the masses as a function 
of the temperature, see the lower two parts in Fig.~\ref{fig:felderli_quarkse}.  
The respective minimum of the sigma mass in the lower left part in Fig.~\ref{fig:felderli_quarkse}   
represents the point of the chiral phase transition. 
From there on the mass of the sigma and the pion start 
to be degenerate.\\ 
For $m_{\sigma}^{vac}=500$~MeV, when neglecting the vacuum term, the sigma and the pion mass come close to the chiral limit. Here $T=130~\rm{MeV}$ and $m_{\sigma}=120~\rm{MeV}$, see also Tab.~\ref{wirsindallebluna1}, and the pion mass nearly jumps vertically around this temperature. 
The inclusion of the vacuum contribution for all values of the initial vacuum mass $m_{\sigma}^{vac}$ leads 
to a less distinctive decrease of $m_{\sigma}$ towards the chiral transition, going along with a clearly less pronounced  minimum, which is also located at 
higher temperatures and higher $m_{\sigma}$ compared to the respective thermal 
value, i.e. when neglecting the vacuum term. 
 From the phase transition point on the mass of the pion, which is seen in the lower right part of Fig.~\ref{fig:felderli_quarkse}, is degenerate to the mass of the sigma. 
 At $T=400$~MeV sigma and pion masses of $\sim$1.2~GeV are achieved. 
\begin{figure}[H]
\center
\includegraphics[width=1.0\columnwidth]{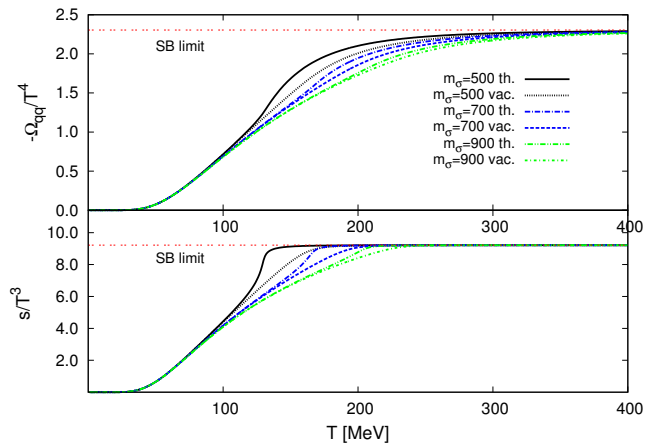}
\caption{\textit{The pressure, divided by $T^4$ as a function of temperature for zero chemical potential without (denoted as ``th.'') and with vacuum contribution (denoted as ``vac.'') for three different values of the initial vacuum sigma meson mass $m_{\sigma}^{vac}$ shown in the upper plot. The  lower plot shows the entropy density s divided by $T^3$ as a function of the temperature. The SB limit represents the Stefan Boltzmann limit.
}}
\label{fig:prentropy_quarkse}
\end{figure}
The upper part in Figure~\ref{fig:prentropy_quarkse} shows the pressure for the three different vacuum sigma meson masses 
including and neglecting the vacuum term. 
All curves rise monotonically. In the temperature region 100 MeV$ \leq T \leq 350$~MeV the curves separate 
and the pressure becomes smaller with increasing value of the vacuum sigma meson mass. 
The inclusion of vacuum fluctuations intensifies this trend at given $m_{\sigma}^{vac}$, so that the pressure 
within this temperature range is smallest for high $m_{\sigma}^{vac}$ and for inclusion of the self energy.
The higher the vacuum mass of the sigma, the less pronounced are the effects from
the inclusion of the vacuum 
fluctuations.
For the smallest value of the 
initial vacuum sigma meson mass $m_{\sigma}^{vac}=500$~MeV and neglecting the vacuum contribution, the 
quarks reach the Stefan Boltzmann limit  (SB limit in the figures) at the lowest temperature, whereas the inclusion of the vacuum contribution at $m_{\sigma}^{vac}=500$~MeV 
pushes down the pressure within the temperature region 100 MeV$ \leq T \leq 350$~MeV. This statement is valid for all $m_{\sigma}^{vac}$, and  can be understood 
as an intrinsic property of the self energy. 
The quarks are more massive for high $m_{\sigma}^{vac}$.
This matches the statement concerning the respective mass spectrum of the sigma and the 
pion at high temperature and can also be observed from the behaviour of the order parameter $\sigma$. 
Recalling that the effective mass of the quarks is generated through the coupling g and the 
fields, see eq.~\ref{eq:sipaen}, this conclusion is not surprising.\\
The lower plot in Figure~\ref{fig:prentropy_quarkse} shows the entropy density 
divided by $T^3$ of the three different initial 
sigma meson masses $m_{\sigma}^{vac}$ including and neglecting the vacuum contributions. 
The entropy density for small $m_{\sigma}^{vac}$ and without the vacuum term  
has higher values at a given temperature compared to the cases with high initial vacuum mass $m_{\sigma}$ 
and the inclusion of the self energy. This feature stems from the fact, that the disorder in the system 
gets larger, the more freely the quarks are. 
Remember, that the higher vacuum 
value $m_{\sigma}^{vac}$, the higher is the temperature, where quarks reach the chiral limit, leading to heavier quarks at intermediate temperatures.
The inclusion of the vacuum energy term amplifies this effect, for low $m_{\sigma}^{vac}$ more 
significantly than for large $m_{\sigma}^{vac}$.
\section{Results for the combined approach} 
At first we neglect the vacuum contribution from the quark- and meson fields, which is denoted as (usual) ``th.`` 
($Q_{th}+M_{th}$ in Tab.~\ref{wirsindallebluna}) 
by setting $F_{dr}(\bar{m}_{\sigma,\vec{\pi}})=0$. 
Even when excluding the mesonic vacuum contribution, the dependence on the quark renormalization scale 
$\Lambda$ does not vanish contrary to the case for the quark fields only, see section \ref{sec:qq_vac}. This is due to 
the contribution from $\Omega_{2PI}$ and corresponds to the case $Q_{th+vac}+M_{th}$ in Tab.~\ref{wirsindallebluna}. 
We choose a value of $\Lambda=1033$~MeV due to reasons which will become clear in 
section \ref{subsectwo}, where we discuss the dependence on both renormalization scales ($Q_{th+vac}+M_{th+vac}$ in Tab.~\ref{wirsindallebluna}).
 \subsubsection{Results for the combined approach 1: Quark vacuum energy}\label{kuarrkselphenergieh}
\begin{figure}[H]
\center
\includegraphics[width=1.0\columnwidth]{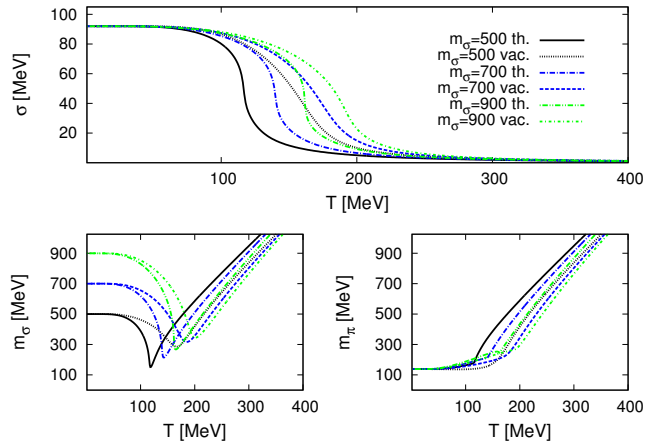}
\caption{\textit{The $\sigma$ condensate in the combined approach as a function of temperature for zero chemical potential without (denoted as ``th.'') and with quark vacuum contribution (denoted as ``vac.'') for three different values of the initial vacuo sigma meson mass $m_{\sigma}^{vac}$ shown in the upper figure. The lower figures show the masses of the sigma and the pion as a function of the temperature. The value of the quark renormalization scale has been chosen to be $\Lambda=1033$~MeV.}}
\label{fig:felderli_qamse}
\end{figure}

\begin{figure}[H]
\center
\includegraphics[width=1.0\columnwidth]{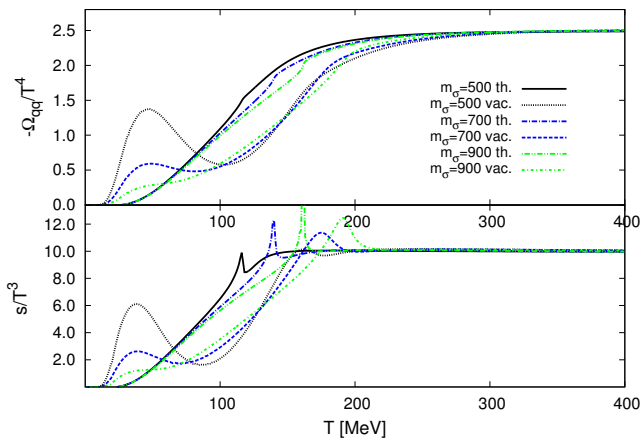}
\caption{\textit{The negative of the potential, i.e. the pressure, divided by $T^4$ as a function of temperature without (denoted as ``th.'') and with vacuum contribution (denoted as ``vac.'') for three different values of the initial vacuo sigma meson mass $m_{\sigma}^{vac}$ shown in the upper figure. The lower figure depicts the entropy density s as a function of the temperature. Some curves show clearly maxima and minima. 
}}
\label{fig:prentropy_qamse}
\end{figure}

The upper figure in  Fig.~\ref{fig:felderli_qamse} shows the order parameter $\sigma$ as 
a function of the temperature within the combined approach for 
the choice of the renormalization scale $\Lambda=1033$~MeV. 
As expected, the larger the value of the initial vacuum sigma meson mass $m_{\sigma}^{vac}$, the further is the 
curve shifted to higher temperatures. The vacuum contribution leads to the same trend as 
when raising the initial value of $m_{\sigma}^{vac}$, so that a high vacuum mass 
$m_{\sigma}^{vac}$ accompanied with the inclusion of the vacuum energy 
leads to the highest phase transition temperature.  
The sigma meson mass as function of the temperature is shown in the lower 
part figure of Fig.~\ref{fig:felderli_qamse}.  
The minima of the sigma meson mass curve, indicating the critical phase transition temperature $T_c$, are closer to the values from the case $Q_{th}$ then from the case $M_{th}$, see Table~\ref{wirsindallebluna1}. This statement is valid in the thermal cases as well when including the fermion vacuum term $Q_{vac}$.  
For low $m_{\sigma}^{vac}$ the minima values are relatively close to the ones from the case $Q_{th}$. Increasing  $m_{\sigma}^{vac}$ shifts the minima, indicating 
that the meson contribution gains influence.\\ 
The behaviour of the pion mass can be seen in the lower right figure in Fig.~\ref{fig:felderli_qamse}. 
The curves seem to be a combination of the pion mass 
spectrum from the case $Q_{th}$ and the 
one from the case $M_{th}$, 
where also the quark contribution dominates. 
For larger values of $m_{\sigma}^{vac}$ the pion mass starts 
to increase at lower temperatures, which is a feature seen for the case $M_{th}$. 
This again underlines the statement that for larger 
sigma meson mass the meson contributions gain influence within the combined approach. In concluding: 
The quarks are dominant in the combined approach. The influence of the meson fields leads to a 
slightly steeper decrease of the order parameter $\sigma$ indicating a trend towards a first order phase transition, 
which is not achieved. 
Both mass spectra in Fig.~\ref{fig:felderli_qamse}  
reach $\sim$1.2 GeV at $T=400$~MeV as is the case for the cases $Q_{th}$ and $Q_{vac}$ exclusively. 
In comparison, the mass spectra in the cases $M_{th}$ and $M_{vac}$ reach $500 \leq m_{\sigma,\pi} \leq 700$~MeV, depending on the initial value of $m_{\sigma}^{vac}$. The vacuum parameters $\lambda$, $m^2$ and $H$, eqs.~(\ref{lambda_qam})-(\ref{H_qam_quasimodo}), for this case $Q_{th+vac}+M_{th}$ are listet in Tab.~\ref{wirsindallebluna}.\\
The pressure of the combined system divided by $T^4$ provided 
by the SU(2) Quark Meson model and the CJT formalism is shown in the upper figure in 
fig.~\ref{fig:prentropy_qamse}.
All curves for the case without the vacuum term start to rise significantly at 
$T\simeq 30$~MeV, whereas the inclusion of the vacuum term causes 
the pressure to rise at $T \simeq 20$~MeV. 
This behaviour results from to the mesonic contributions.
The curves show distinct extrema, 
less pronounced with larger $m_{\sigma}^{vac}$, located around $T\simeq 45$~MeV. 
This clearly is correlated to the influence of the vacuum term 
leading to a higher pressure at given temperature compared to the case without 
the vacuum term.
In the combined approach this leads to distinct extrema, indicating the dominance 
of the meson contribution at low temperature. It is important 
to note that these extrema are not instabilities, since the pressure 
itself is a monotonically rising function, and so is the entropy density, which is seen in the lower figure in fig. \ref{fig:prentropy_qamse}. 
Neglecting the vacuum contribution, the curves also exhibit a nontrivial behaviour within 
the temperature range $100 \leq T \leq 180$~MeV, again leading to very distinctive maxima in 
the entropy density.
The entropy density curves without vacuum term rise 
approximately linear at low temperature. For $m_{\sigma}^{vac}=500$~MeV 
a maximum at $T=116$~MeV and $s/T^3=9.85$ can be observed, which can be 
traced back to the hardly visible change of slope 
in the pressure in the upper figure.
The higher the vacuum sigma meson mass, the more 
pronounced are the maxima in $s/T^3$. This occurs in all 
cases considered at the phase transition. These peaks arise from the fact that the pressure has a considerably 
change of slope at the chiral phase transition temperature.
A possible explanation of having two maxima might be that 
the change of the relativistic degrees of freedom occurs in two different temperature regions. 
One can interpret these pronounced peaks as an intermediate 
sudden increase in relativistic degrees of freedom or as an field 
energy contribution. 
Note also, that an entropy jump as in a first 
order phase transition is not observed.\\ 
%
\begin{table}
\begin{center}
\begin{tabular}{|c||c|c||c|c||c|c||}
\hline\hline 
\multicolumn{1}{|c||}{$m_{\sigma}^{vac}$} & \multicolumn{2}{|c||}{$Q_{th/vac}$} & \multicolumn{2}{c||}{$M_{th/vac}$}  & \multicolumn{2}{c||}{$Q_{th/vac}+M_{th}$}\\
\hline
 & \hspace{.2cm} T \hspace{.2cm} & \hspace{.1cm} $m_{\sigma}$ \hspace{.1cm} & \hspace{.2cm} T \hspace{.2cm} & \hspace{.1cm} $m_{\sigma}$ \hspace{.1cm}& \hspace{.2cm} T \hspace{.2cm} & $m_{\sigma}$ \\
\cline{2-7}
\cline{2-7}
$500_{(th)}$ & 130 & 120 & 230 & 290 & 118 & 150   \\
$500_{(th+vac)}$ & 163 & 287 & 260 & 320 &  166 & 285  \\ 
 \cline{1-7}
$700_{(th)}$ & 165 & 185 & 238 & 324 &  143 & 214  \\ 
$700_{(th+vac)}$ & 198 & 310 & 305 & 414 &  185 & 316  \\
\cline{1-7}
$900_{(th)}$  & 205 & 243 & 245 & 355 &  165 & 267   \\
$900_{(th+vac)}$  & 233 & 336 & 360 & 510 &  201 & 344   
\\ \hline\hline
\end{tabular}
\caption{\textit{The minimal mass for the $\sigma$-meson for all three different approaches, i.e. quarks with and without vacuum term, case $Q_{th/vac}$,   (section~\ref{qq_i_actins}), mesons with and without vacuum term, case $M_{th/vac}$, (section~\ref{mesonic_inter}) and quarks and mesons combined with and without vacuum term for the quark fields, $Q_{th/vac}+M_{th}$, (section~\ref{ciwqam}). All values are given in MeV.}}
\label{wirsindallebluna1}
\end{center}
\end{table}

Tab.~\ref{wirsindallebluna1} shows the minimal value 
of the sigma meson mass in the medium for the cases $Q_{th}$, $Q_{vac}$, $M_{th}$, $M_{vac}$ and for $Q_{th/vac}+M_{th}$. 
With or without the vacuum term 
the minima of the combined approach are closer to the values of the 
thermal quarks then to the values for thermal mesons.
The impact of the thermal mesons shifts the minima of the combined approach 
to lower temperatures. 
 \subsubsection{Results for the combined approach 2: Dependence on the renormalization scale}\label{subsectwo}
In this section, we explore the impact of having two renormalization scales, 
one from the quark fields $\Lambda$
and one from the mesonic fields $\mu$. This corresponds to the case $Q_{th+vac}+M_{th+vac}$ in Tab.~\ref{wirsindallebluna}. 
In the last subsection we set $F_{dr}(\bar{m}_{\sigma,\vec{\pi}})=0$, omitting the 
self energy resulting from the 2PI formalism from the mesonic fields. 
In this section we show that this contribution is negligible 
for the fields and the mass spectra, but not for the thermodynamics, 
i.e. the respective relativistic degrees of freedom. 
First we run the code with one value for the 
renormalization scale, i.e. setting $\Lambda=\mu$ and in a second approach we keep $\mu$ fixed at the 
value used in \cite{Grahl:2011yk}, that is $\mu=m_{\sigma}/\sqrt{e}$. 
We first study the three vacuum parameters $\lambda$ (eq.~(\ref{lambda_qam})), 
$m^2$ (eq.~(\ref{mquadraht_qam})) and $H$ (eq.~(\ref{H_qam_quasimodo})) 
as a function of the renormalization scale for $\Lambda=\mu$ 
and for the choice $\mu=m_{\sigma}/\sqrt{e}$, 
such as to locate the most reasonable renormalization scale value, 
which turns out to be the one used in the previous section, $\Lambda=1033$~MeV. 
The value of the sigma meson mass has been chosen to be at a value of $m_{\sigma}=550$~MeV. 
The renormalization scale parameter is naturally placed at 
the chiral scale \cite{Lenaghan:1999si,Lenaghan:2000ey,Mocsy:2004ab}, 
i.e. is of the order 1 GeV. Setting $\Lambda=\mu$ or even $\mu=m_{\sigma}/\sqrt{e}$ 
we find reasonable solutions only within the range 
$850 \leq \Lambda \leq 1150$~MeV, which we investigate during this section.\\

\begin{figure}[H]
\center
\includegraphics[width=1.0\columnwidth]{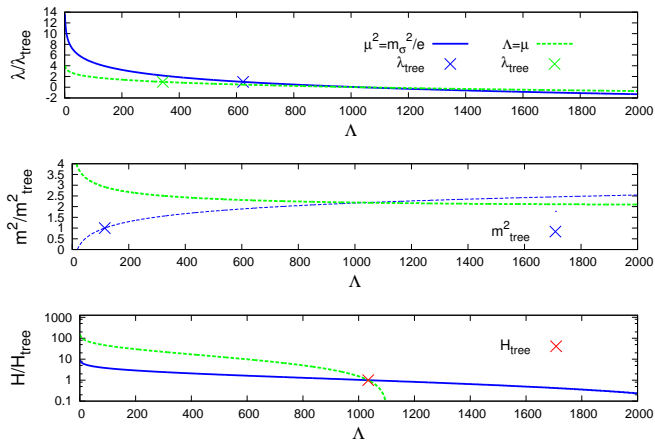}
\caption{\textit{The vacuum parameters $\lambda$, $m^2$ and H, normalized to their respective tree level value ($\lambda\simeq16.64$, $m^2=-122683\rm{MeV}^2$ and $H=1.75\cdot10^6\rm{MeV}^3$), as a function of the quark renormalization scale $\Lambda$. The cross marks the tree level values.
}}
\label{fig:ren_scales}
\end{figure}

\begin{figure}[H]
\center
\includegraphics[width=1.0\columnwidth]{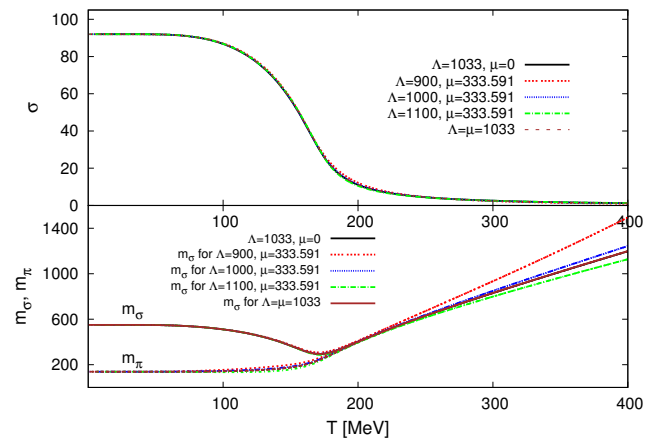}
\caption{\textit{The $\sigma$ condensate as a function of temperature for the renormalization scales $\Lambda=1033$~MeV and $\mu=0$ (Sec.~\ref{kuarrkselphenergieh}), $\Lambda=900,1000,1100$~MeV at $\mu=m_{\sigma}/\sqrt{e}$ held fixed and for $\Lambda=\mu=1033$~MeV at $m_{\sigma}=550$~MeV shown in the upper figure. The lower figure shows the sigma and the pion mass spectrum.}}
\label{fig:f_a_18}
\end{figure}

Fig.~\ref{fig:ren_scales} 
shows the coupling $\lambda$, the mass term $m^2$ and 
the explicit symmetry breaking term $H$ normalized to their respective tree level values as a function 
of the renormalization scale with $\Lambda=\mu$ (dotted curve) 
and with $\mu=m_{\sigma}/\sqrt{e}=333.591$~MeV held fixed (continuous curve). The respective values are also given in  
Tab.~\ref{wirsindallebluna}.\\ 
The tree level value for $\lambda$ for  
the choice $\Lambda=\mu$ is found to be located at $\Lambda=343$~MeV, 
which is surprisingly close to $\mu=m_{\sigma}/\sqrt{e}$~MeV. However,  
for the choice $\mu=m_{\sigma}/\sqrt{e}$~MeV the tree level 
value is located at $\Lambda=623$~MeV.
Note that the two curves in the upper figure intersect at $\Lambda=1033$~MeV.\\ 
The tree level value of $m^2$ for $\Lambda=\mu$ is never reached (middle figure), and 
when setting 
$\mu=m_{\sigma}/\sqrt{e}$~MeV the curve surprisingly increases with $\Lambda$, and the tree level 
value is located at $\Lambda=115$~MeV. 
These two curves also intersect at $\Lambda=1033$~MeV.\\
The explicit symmetry breaking term $H$, which is responsible for the 
mass of the pion, is shown normalized to its tree level value in the 
lowest figure in Fig.~\ref{fig:ren_scales}.
The tree level value is for both choices ($\Lambda=\mu$ and for $\mu=m_{\sigma}/\sqrt{e}$) located at $\Lambda=1033$~MeV, where these two curves also intersect 
(which motivates our choice for $\Lambda=\mu=1033$~MeV in the previous subsection).\\
The order parameter $\sigma$ for different renormalization 
scales is shown in the upper part in Fig.~\ref{fig:f_a_18}, whereas the lower 
part shows the mass spectrum of the sigma and the pion.\\ 

Fig. \ref{fig:f_a_18} contains the calculation for only one renormalization scale with $\Lambda=1033$~MeV and $\mu=0$ for $m_{\sigma}^{vac}=550$~MeV from Sec.~\ref{kuarrkselphenergieh} for comparison. 
For the choice for $\mu$ according to 
\cite{Grahl:2011yk} we choose three values of $\Lambda$ and finally we set 
$\Lambda=\mu=1033$~MeV.
All cases show a crossover phase transition at $T\simeq 165$~MeV, and there is no notable difference in the order parameter. 
The different cases for the mass spectrum do not show significant differences up to 
$T\simeq250$~MeV, where the degenerate masses of the sigma and the pion start to have different slopes. 
It is worth mentioning that the curves are very similar 
to the curves from case $Q_{th}$ or $Q_{vac}$ and result in similar masses at large temperatures, 
demonstrating again the dominance of the quark contribution.\\

\begin{figure}[H]
\center
\includegraphics[width=1.0\columnwidth]{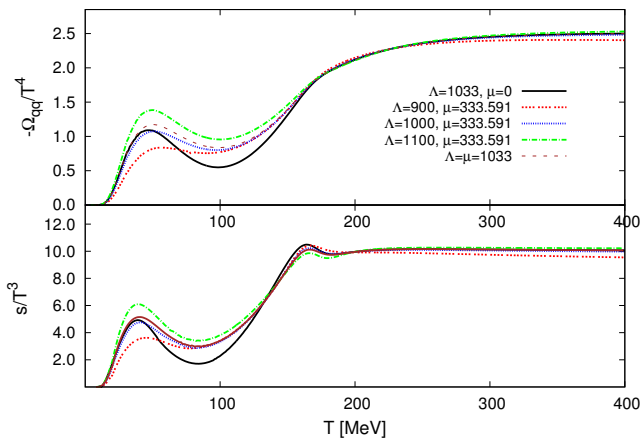}
\caption{\textit{The pressure, divided by $T^4$ as a function of of temperature for the renormalization scale choices $\Lambda=1033$~MeV and $\mu=0$ (Sec.~\ref{kuarrkselphenergieh}), $\Lambda=900,1000,1100$~MeV at $\mu=m_{\sigma}/\sqrt{e}$ held fixed and for $\Lambda=\mu=1033$~MeV at $m_{\sigma}=550$~MeV is shown in the upper figure. The lower figure shows the entropy density s divided by $T^3$ as a function of the temperature.}}
\label{fig:f_a_20}
\end{figure}

The pressure divided by $T^4$ as a function of of temperature 
for the renormalization scale choices $\Lambda=1033$~MeV and $\mu=0$ (Sec.~\ref{kuarrkselphenergieh}), $\Lambda=900,1000,1100$~MeV 
at $\mu=m_{\sigma}/\sqrt{e}$ held fixed and for $\Lambda=\mu=1033$~MeV 
at $m_{\sigma}=550$~MeV are represented in the upper figure in Fig.~\ref{fig:f_a_20}.
All the curves show two maxima, 
one at $T \simeq 50$~MeV and a smaller one around the phase transition at $T \simeq 165$~MeV. 
For $\Lambda=1033$~MeV and $\mu=0$ the maximum is located within the same region as for two renormalization scales, whereas the minimum is shifted to a considerably lower value of $p/T^4$. 
The second extrema are a result from the contribution from the mesonic fields and are changing slightly with the choice of the renormalization scale.  


%

%
\begin{table*}
\begin{center}
\begin{tabular}{|c|c|c|c|c|c|c|}
\hline\hline 
\textbf{Case} & $m_{\sigma}^{vac}$ & $\Lambda$ & $\mu$  & $\lambda$ & $m^2$  & $H$ \\
\hline
$Q_{th}$ & 500 & - & - & 16.744 & -122683 & $1.75\cdot 10^{6}$  \\
\cline{2-6}
$Q_{th+vac}$ & 500 & - & - & 42.521 & -268130 & $1.75\cdot 10^{6}$  \\
\cline{2-6}
$M_{th}$ & 500 & - & - & 16.744 & -122683 &  $1.75\cdot 10^{6}$  \\ 
\cline{2-6}
$M_{th+vac}$ & 500 & - & 333.591 & 16.11 & -90449 &  $2.74\cdot 10^{6}$  \\ 
\cline{2-6}
$Q_{th}+M_{th}$ & 500 & - & - & 16.744 & -122683 &  $1.75\cdot 10^{6}$  \\ 
\cline{2-6}
$Q_{th+vac}+M_{th}$ & 550 & 1033 & - & 0.0268 & -268130 &  $1.75\cdot 10^{6}$  \\ 
\cline{2-6}
$Q_{th+vac}+M_{th+vac}: \Lambda=\mu$ & 550 & 1033 & 1033 & 0.013 & -268148 &  $1.77\cdot 10^{6}$  \\ 
\cline{2-6}
$Q_{th+vac}+M_{th+vac}: \Lambda \neq \mu$ & 550 & 900 & 333.591 & 4.583 & -258959 &  $2.03\cdot 10^{6}$  \\ 
\cline{2-6}
$Q_{th+vac}+M_{th+vac}: \Lambda \neq \mu$ & 550 & 1000 & 333.591 & 1.099 & -265930 &  $1.82\cdot 10^{6}$  \\ 
\cline{2-6}
$Q_{th+vac}+M_{th+vac}: \Lambda \neq \mu$ & 550 & 1100 & 333.591 & -2.052 & -272236 &  $1.62\cdot 10^{6}$  \\ \hline\hline
\end{tabular}
\caption{\textit{The parameters $\lambda$, $m^2$ and $H$ for all considered cases. Thermal quarks are labeled $Q_{th}$, including the vacuum term for the quark fields is labeled $Q_{th+vac}$. Thermal mesons without vacuum term are labeled $M_{th}$ and with vacuum term $M_{th+vac}$. The approach combining quarks and mesons a without vacuum term is labeled $Q_{th}+M_{th}$. For these cases the sigma meson mass is $m_{\sigma}^{vac}=500$~MeV. The combination of both approaches with vacuum term only for the quark fields is labeled $Q_{th+vac}+M_{th}$, and with vacuum term in both approaches $Q_{th+vac}+M_{th+vac}$. Here $m_{\sigma}^{vac}=550$~MeV for the different choices of the renormalization scale, which is given in MeV. $\lambda$ is dimensionless, $m^2$ in $MeV^2$ and H is given in $MeV^3$.}}
\label{wirsindallebluna}
\end{center}
\end{table*}

  \section{Conclusions}
  
  
In this article we have studied quarks, with the common path integral formalism, and mesons, utilizing the 2PI formalism, within the SU(2) Quark Meson model at zero chemical potential in a combined set of equations.

We investigated the influence of the vacuum fluctuations for different values of the sigma meson mass and for different choices of the renormalization scale parameters on the order parameter, the mass spectra of the sigma and the pion and for thermodynamical quantities.\\ 
The inclusion of the vacuum fluctuations for the quark fields 
is independent of the renormalization scale \cite{Gupta:2011ez,Chatterjee:2011jd},  
whereas for the meson fields the dependence on the renormalization 
scale does not cancel. Inclusion of the vacuum term for the quark fields leads to a distinct shift of the chiral phase transition to higher  temperatures. 
The inclusion of the vacuum contribution turn out to be in both cases not negligible.
Within the combined case 
we were hence left with the option of having two renormalization scales or one for quarks and mesons. \\
We investigated separately the vacuum parameters $\lambda$, $m^2$ and $H$ as 
a function of the quark renormalization scale $\Lambda$ and conclude that 
the main impact comes from the quark fields. There is a tiny window around $\Lambda \sim$~1 GeV, where the results are physically reasonable, i.e. close to tree-level values.
The fields and the mass spectra showed hardly any difference 
when varying the renormalization scale.  
It seems that the thermal contribution of the mesons have an influence within 
the temperature region $50 \leq T \leq 180$~MeV for the pressure, which gives rise to 
peaks within the entropy to temperature ratio. According to lattice QCD calculations, this behaviour is 
clearly unphysical \cite{Borsanyi:2010bp}, so that only the results for $Q_{th+vac}$ and $Q_{th}+M_{th}$ are employable.   
We find that in all cases considered a chiral first order phase transition is 
not present.\\ 
Ref.~\cite{Scavenius:2000qd} compares the renormalized linear sigma model with the 
NJL model. Like in our case a crossover transition has been found for zero chemical potential 
and The authors stress the importance of the vacuum field 
fluctuations to the thermodynamic properties. 
In Ref.~\cite{Mocsy:2004ab} the linear sigma model including the vacuum field fluctuations, containing 
quark and mesonic degrees of freedom, 
has been studied. The quark degrees of freedom have been integrated 
out and the resulting effective action was linearized around the ground 
state. Sigma mesons and pions were described as 
quasiparticles and their properties were taken into account within 
the thermodynamic potential. Their parameter choice is similar to ours and 
they find a gradual decrease of the chiral condensate, which results   
in a crossover type transition at temperatures 
$150 \leq T_c \leq 200$~MeV. Also the results for the masses are very similar 
to our results. 
Their thermodynamical quantities do not show such an influence from the meson fields  
in the low temperature region.  
We argue that this feature comes from the 2PI formalism used in our work.\\
Future work could implement the Polyakov loop to mimic the quark confinement 
\cite{Sasaki:2006ww,Schaefer:2007pw,Stiele:2013pma,Stiele:2016cfs}. It would also be 
interesting to perform calculations for non-zero chemical potential to explore the 
QCD phase diagram \cite{Herbst:2013ufa} or calculations for finite isospin \cite{Stiele:2013pma}. 
The implementation of the strange quark \cite{Karsch:2000kv,Karsch:2000hh} in a SU(3) Quark Meson model, and, if 
applicable, vector mesons \cite{Carter:1996rf,Carter:1995zi}, could yield a realistic model for 
astrophysical applications, such as for proto neutron stars or neutron star merger \cite{Heinimann:2016zbx,Hanauske:2016gia}. In \cite{Zacchi:2015lwa,Zacchi:2015oma,Zacchi:2016tjw} we have already shown that the SU(3) approach in the mean field approximation yields realistic compact star scenarios. Hence the expansion of the SU(3) quark meson model to finite temperatures with the vacuum term or a combined approach with quark- and meson fields in the mean field approximation could indeed yield an appropriate model for a quark based equation of state for astrophysical application. 
\begin{acknowledgments}
The authors thank Dirk Rischke, Rainer Stiele and Thorben Graf 
for discussions during the initial stage 
of this project. Furthermore we want to thank Konrad Tywoniuk 
(CERN) for helpful suggestions concerning the renormalization process.
AZ is supported by the Stiftung Giersch.
\end{acknowledgments}
\bibliography{neue_bib}
\bibliographystyle{apsrev4-1}
\end{document}